\documentclass[twocolumn,showpacs,prl]{revtex4}
\usepackage{amssymb} 
\usepackage{amsmath}
\usepackage{graphicx}
\usepackage{dcolumn}
\usepackage{bm}

\begin{document}

\title{Entanglement detection in the vicinity of arbitrary Dicke states}
\author{L.-M. Duan$^{1,2}$}
\address{$^{1}$Department of Physics and MCTP, University of Michigan, Ann
Arbor, Michigan 48109, USA}
\address{$^{2}$Center for Quantum Information, IIIS, Tsinghua University,
Beijing, China}

\begin{abstract}
Dicke states represent a class of multipartite entangled states that can be
generated experimentally with many applications in quantum information. We
propose a method to experimentally detect genuine multipartite entanglement
in the vicinity of arbitrary Dicke states. The detection scheme can be used
to experimentally quantity the entanglement depth of many-body systems and
is easy to implement as it requires to measure only three collective spin
operators. The detection criterion is strong as it heralds multipartite
entanglement even in cases where the state fidelity goes down exponentially
with the number of qubits.
\end{abstract}

\maketitle

Quantum entanglement provides the most useful resource for implementation of
many quantum information protocols. To test fundamentals of quantum
mechanics and to realize quantum information processing, a big experimental
drive is to get more and more particles prepared into massively entangled
states \cite{1,2,3,4}. There are different types of entangled states for
many qubits \cite{5,6,7}. Experiments so far typically center around two
kinds of entangled states \cite{1,2,3,4}. The first kind is the graph
states, including the GHZ\ states as a special case \cite{1}; the second
kind is the Dicke states, including the W states as a special case \cite%
{2,3,4}. Both types of entangled states have interesting properties and
important applications in quantum information \cite{5,6,7}, and they have
been generated from a number of experimental systems \cite{1,2,3,4}. One can
never get a perfect entangled state in any experiments. A critical question
is thus to experimentally prove that the prepared state still contains
genuine multipartite entanglement similar to the target state. For graph
states, some powerful witness operators have been known which significantly
simplifies the experimental entanglement detection \cite{1,8,9}. For Dicke
type of states, however, the entanglement detection is more challenging. The
experiments so far use either quantum states tomography \cite{2}, which
requires measurements in an exponentially large number of experimental
settings and thus is limited to only small systems, or some clever tricks
that apply to only particular Dicke states \cite{3,4,10,11}, and are hard to
be generalized to arbitrary Dicke states of many qubits.

In this paper, we propose a general method to detect genuine multipartite
entanglement in the vicinity of arbitrary Dicke states and to characterize
the entanglement depth of the system. The proposed scheme has the following
favorable features: first, it only requires to measure the collective spin
operators and thus is straightforward for experimental implementation.
Independent of the number of qubits, we only need to measure three operators
with no requirement of separate addressing of individual qubits. This is
particularly convenient for entanglement detection in many-particle systems
(such as a spinor condensate) where individual addressing is almost
impossible. Second, the proposed detection criterion is strong and
universally applicable to arbitrary Dicke states. It not only detects
entanglement, but also quantifies the entanglement depth of the system \cite%
{note,13}. The detection scheme is pretty robust to experimental noise, and
can show significant entanglement depth of the system even in cases where
the state fidelity has been exponentially small with the number of qubits.

The Dicke states are co-eigenstates of the collective spin operators. Each
qubit is described by a Paul matrix $\mathbf{\sigma }$. For $N$ qubits, we
define the collective spin operator $\mathbf{J}$ as $\mathbf{J=}%
\sum_{i=1}^{N}\mathbf{\sigma /}2$. The Dicke state $\left\vert
N/2,n/2\right\rangle $ is defined as a coeigenstate of the operators $%
\mathbf{J}^{2}\equiv J_{x}^{2}+J_{y}^{2}+J_{z}^{2}$ and $J_{z}$, with the
eigenvalues $N(N+2)/4$ and $n/2$ ($n=-N/2,-N/2+1,\cdots ,N/2$),
respectively. The Dicke states can be conveniently generated in experiments
without the need of separate addressing \cite{5,6,7,14}. Except for the
trivial case of $n=\pm N$, the Dicke states is a multipartite entangled
state with interesting applications in both precision measurements and
quantum information \cite{2,3,4,5,14,15}.

To construct an entanglement detection criterion in the vicinity of Dicke
states, we note that the variances of the collective spin operators $%
J_{x},J_{y},J_{z}$ have very special properties for these states. The
variance of $J_{z}$ is minimized (ideally it should be zero) while the
variances of $J_{x},J_{y}$ are maximized under the constraint of $%
\left\langle J_{z}\right\rangle $. So, to detect entanglement, we should
construct an inequality to bound the variances of $J_{x},J_{y}$ with the
variance of $J_{z}$ for any separable states or insufficiently entangled
states, and at the same time this inequality should be violated by the
states sufficiently close to a Dicke state.

For a composite system of $N$ qubits, we note that its density operator $%
\rho $ can always be written into the following form if $\rho $ does not
contain genuine $N$-qubit entanglement \cite{16}:
\begin{equation}
\rho =\sum_{\mu }p_{\mu }\rho _{\mu },  \label{1}
\end{equation}%
with $p_{\mu }\geq 0$, $\sum_{\mu }p_{\mu }=1$, and
\begin{equation}
\rho _{\mu }=\rho _{1\mu }\otimes \rho _{2\mu }\otimes \cdots \otimes \rho
_{k\mu },  \label{2}
\end{equation}%
where $\rho _{i\mu }$ ($i=1,2,\cdots ,k$) represents a component state of $%
m_{i\mu }$ ($m_{i\mu }\geq 1$) qubits with $\sum_{i=1}^{k}m_{i\mu }=N$. In
other words, for each component $\mu $, the $N$ qubits are divided into $k$
groups with $m_{i\mu }$ qubits for the $i$th group, and the component state $%
\rho _{\mu }$ is a tensor product of the states for each group. For a fixed
component $\mu $, each qubit uniquely belongs to one group, however, for
different $\mu $, the group division of the qubits can be different. If all $%
m_{i\mu }=1$ (and corresponding $k=N$), $\rho $ reduces to a separable
state. If the maximum of $m_{i\mu }$ is $m_{0}$, we conclude that the state $%
\rho $ has no genuine ($m_{0}+1$)-qubit entanglement \cite{16}. With a
smaller $m_{0}$, the state $\rho $ gets less entangled.

We now show that for any states in the form Eqs. (1-2), the variance of the
collective spin operators are severely bounded, while this bound is violated
by the Dicke states. For each group division $\mu $ of the $N$ qubits, the
total collective spin operators $\mathbf{J}$ can be written as $\mathbf{J}%
=\sum_{i=1}^{k}\mathbf{J}_{i}$, where $\mathbf{J}_{i}=\sum_{j=1}^{m_{i\mu }}%
\mathbf{\sigma }_{j}/2$ is the collective spin operator for $m_{i\mu }$
qubits in the $i$th group. Through addition of the angular momenta, we know
the maximum spin of $\mathbf{J}_{i}$ is $m_{i\mu }/2$, so the moments of $%
J_{\alpha i}$ $\left( \alpha =x,y,z\right) $\ are bounded by
\begin{equation}
\left\langle J_{\alpha i}^{2}\right\rangle \leq m_{i\mu }^{2}/4,\text{and }%
\left\langle \mathbf{J}_{i}^{2}\right\rangle \leq m_{i\mu }(m_{i\mu }+2)/4.
\label{3}
\end{equation}%
Under state $\rho $, we have $\left\langle J_{x}^{2}\right\rangle =\sum_{\mu
}p_{\mu }\left\langle J_{x}^{2}\right\rangle _{\mu }$ and%
\begin{equation}
\left\langle J_{x}^{2}\right\rangle _{\mu }=\sum_{i_{1},i_{2}}\left\langle
J_{xi_{1}}\right\rangle _{\mu }\left\langle J_{xi_{2}}\right\rangle _{\mu
}+\sum_{i}\left\langle \left( \Delta J_{xi}\right) ^{2}\right\rangle _{\mu }.
\label{4'}
\end{equation}%
Using the uncertainty relation $\left\langle \left( \Delta J_{yi}\right)
^{2}\right\rangle _{\mu }\left\langle \left( \Delta J_{zi}\right)
^{2}\right\rangle _{\mu }\geq \left\langle J_{xi}\right\rangle _{\mu
}^{2}/4, $we can bound the term $\sum_{i_{1},i_{2}}\left\langle
J_{xi_{1}}\right\rangle _{\mu }\left\langle J_{xi_{2}}\right\rangle _{\mu }$
as
\begin{widetext}

\begin{eqnarray}
&&\sum_{i_{1},i_{2}}\left\langle J_{xi_{1}}\right\rangle _{\mu }\left\langle
J_{xi_{2}}\right\rangle _{\mu } \leq \sum_{i_{1},i_{2}}4\sqrt{\left\langle
\left( \Delta J_{yi_{1}}\right) ^{2}\right\rangle _{\mu }\left\langle \left(
\Delta J_{zi_{1}}\right) ^{2}\right\rangle _{\mu }\left\langle \left( \Delta
J_{yi_{2}}\right) ^{2}\right\rangle _{\mu }\left\langle \left( \Delta
J_{zi_{2}}\right) ^{2}\right\rangle _{\mu }}  \notag \\
&\leq &\sum_{i_{1},i_{2}}2\left[ \left\langle \left( \Delta
J_{yi_{1}}\right) ^{2}\right\rangle _{\mu }\left\langle \left( \Delta
J_{zi_{2}}\right) ^{2}\right\rangle _{\mu }+\left\langle \left( \Delta
J_{yi_{2}}\right) ^{2}\right\rangle _{\mu }\left\langle \left( \Delta
J_{zi_{1}}\right) ^{2}\right\rangle _{\mu }\right] =4\left\langle \left(
\Delta J_{z}\right) ^{2}\right\rangle _{\mu }\sum_{i}\left\langle \left(
\Delta J_{yi}\right) ^{2}\right\rangle _{\mu },
\end{eqnarray}%

\end{widetext}

where we have used the relation $\left\langle \left( \Delta J_{z}\right)
^{2}\right\rangle _{\mu }=\sum_{i}\left\langle \left( \Delta J_{zi}\right)
^{2}\right\rangle _{\mu }$ for the state in the form of Eqs. (1-2).
Combining Eqs. (4) and (5), we get
\begin{equation}
\left\langle J_{x}^{2}\right\rangle \leq \sum_{\mu ,i}p_{\mu }\left[
\left\langle \left( \Delta J_{xi}\right) ^{2}\right\rangle _{\mu
}+4\left\langle \left( \Delta J_{z}\right) ^{2}\right\rangle _{\mu
}\left\langle \left( \Delta J_{yi}\right) ^{2}\right\rangle _{\mu }\right] .
\label{6'}
\end{equation}%
Using the relation $\left\langle \left( \Delta J_{\alpha i}\right)
^{2}\right\rangle _{\mu }\leq \left\langle J_{\alpha i}^{2}\right\rangle
_{\mu }\leq m_{i\mu }^{2}/4$ (see Eq. (3)) and $\left\langle \left( \Delta
J_{z}\right) ^{2}\right\rangle \geq \sum_{\mu }p_{\mu }\left\langle \left(
\Delta J_{z}\right) ^{2}\right\rangle _{\mu }$, we can bound $\left\langle
J_{x}^{2}\right\rangle $ by
\begin{equation}
\left\langle J_{x}^{2}\right\rangle \leq \left[ 1+4\left\langle \left(
\Delta J_{z}\right) ^{2}\right\rangle \right] \max_{\left\{ m_{i\mu
}\right\} }\left( \sum_{i=1}^{k_{u}}m_{i\mu }^{2}/4\right) ,  \label{7'}
\end{equation}%
where the maximum is taken over all the group division $\left\{ m_{i\mu
}\right\} $ ($m_{i\mu }$ are positive integers) of the $N$ qubits with the
constraint of $\sum_{i=1}^{k}m_{i\mu }=N$ and $m_{i\mu }\leq m_{0}$. The
maximum value is obtained by choosing $k=\left\lceil N/m_{0}\right\rceil $ ($%
\left\lceil N/m_{0}\right\rceil $ denotes the smallest integer no less than $%
N/m_{0}$), $m_{1\mu }=N-m_{0}(k-1)$, and all the other $m_{i\mu }=m_{0}$ ($%
i=2,\cdots ,k$). Correspondingly, Eq. (10) reduces to%
\begin{equation}
\left\langle J_{x}^{2}\right\rangle \leq \left[ 1+4\left\langle \left(
\Delta J_{z}\right) ^{2}\right\rangle \right] m_{0}N/4,  \label{8'}
\end{equation}%
where we have used the relation $m_{1\mu }^{2}+m_{0}^{2}\left( k-1\right)
\leq m_{0}\left[ m_{1\mu }+m_{0}\left( k-1\right) \right] =m_{0}N$. So, for
any states without genuine ($m_{0}+1$)-qubit entanglement, the moment $%
\left\langle J_{x}^{2}\right\rangle $ (and similarly also $\left\langle
J_{y}^{2}\right\rangle $) will be bounded by the inequality (8). When $%
m_{0}\geq 2$, we can derive a stronger bound. Note that $\left\langle
J_{y}^{2}\right\rangle $ satisfies an inequality similar to Eq. (6), but
with the indices $x$ and $y$ exchanged. If we add up the inequalities for $%
\left\langle J_{x}^{2}\right\rangle $ and $\left\langle
J_{y}^{2}\right\rangle $, and use the relation $\left\langle \left( \Delta
J_{xi}\right) ^{2}\right\rangle _{\mu }+\left\langle \left( \Delta
J_{yi}\right) ^{2}\right\rangle _{\mu }\leq \left\langle \mathbf{J}%
_{i}^{2}\right\rangle \leq m_{i\mu }(m_{i\mu }+2)/4$ (see Eq. (3)), we
obtain
\begin{equation}
\left\langle J_{x}^{2}\right\rangle +\left\langle J_{y}^{2}\right\rangle
\leq \left[ 1+4\left\langle \left( \Delta J_{z}\right) ^{2}\right\rangle %
\right] N\left( m_{0}+2\right) /4.  \label{9'}
\end{equation}

We can use violation of the inequality (8) with $m_{0}=1$ to experimentally
prove entanglement of the system and then use the following criterion to
quantity its entanglement depth:

\textit{Criterion 1: We can experimentally measure the following quantity }$%
\xi $\textit{\ through detection of the collective spin operator }$\mathbf{J}
$\textit{:}%
\begin{equation}
\xi =\frac{\left\langle J_{x}^{2}\right\rangle +\left\langle
J_{y}^{2}\right\rangle }{N\left( 1/4+\left\langle \left( \Delta J_{z}\right)
^{2}\right\rangle \right) }-1.  \label{10'}
\end{equation}%
\textit{If }$\xi >m$\textit{, it is confirmed that the system has genuine }$%
m $\textit{-qubit entanglement. }

For the Dicke state $\left\vert N/2,0\right\rangle $, we have $\left\langle
J_{x}^{2}\right\rangle =\left\langle J_{y}^{2}\right\rangle =N\left(
N+2\right) /8$ and $\left\langle \left( \Delta J_{z}\right)
^{2}\right\rangle =0$, so in the ideal case, $\xi =N+1>N$, and from
measurement of $\xi $, we can confirm that all the qubits are in a genuine $N
$-qubit entangled state. The noise in experiments will degrade the
entanglement depth of the system. First, we consider dephasing noise which
is a major source of noise in many experiments. The detection criterion in
Eq. (10) is very robust to dephasing noise. To see this, we note the state $%
\left\vert N/2,0\right\rangle $ is a big superposition state with $\binom{N}{%
N/2}=\frac{N!}{\left[ \left( N/2\right) !\right] ^{2}}$ terms in the
computational basis. All the superpositions terms have $J_{z}=0$, so the
dephasing error only degrades the moments $\left\langle
J_{x}^{2}\right\rangle +\left\langle J_{y}^{2}\right\rangle $, but does not
increase $\left\langle \left( \Delta J_{z}\right) ^{2}\right\rangle $. For
each superposition term of the state $\left\vert N/2,0\right\rangle $, we
know $\left\langle J_{y}^{2}\right\rangle =\left\langle
J_{x}^{2}\right\rangle =\sum_{i=1}^{N}\left\langle \left( \sigma
_{ix}/2\right) ^{2}\right\rangle =N/4$. So, if coherence is completely gone,
$\xi $ reduces to $1$, and the state has no entanglement as expected.
However, under incomplete dephasing, we can experimentally prove a
significant entanglement depth of the system by measuring $\xi $ even if the
state fidelity becomes exponentially small. For instance, with a dephasing
error rate $p$ for each qubit, the state fidelity goes down exponentially
roughly by $p^{N}$ for $N$ qubits with $N\gg 1$. To estimate the value of $%
\xi $, we note that with a probability $\binom{N}{i}p^{i}\left( 1-p\right)
^{N-i}$ (according to the binormal distribution), $i$ qubits are decohered
among the $N$ qubits, which contribute a value of $i/2$ to $\left\langle
J_{x}^{2}\right\rangle +\left\langle J_{y}^{2}\right\rangle $. The remaining
$N-i$ qubits still have coherence, which contribute a value of $\left(
N-i\right) \left( N-i+2\right) /4-\left\langle J_{z}^{2}\right\rangle _{N-i}$
to $\left\langle J_{x}^{2}\right\rangle +\left\langle J_{y}^{2}\right\rangle
$. Since initially the $N$\ qubits are in the $J_{z}=0$ eigenstate, the mean
value of $\left\langle J_{z}^{2}\right\rangle _{N-i}$ for the $N-i$ qubits
is equal to $\left\langle J_{z}^{2}\right\rangle _{i}$ for the decohered $i$
qubits. For the decohered $i$ qubits, $\left\langle J_{z}^{2}\right\rangle
_{i}=\sum_{k=1}^{i}\left\langle \left( \sigma _{kz}/2\right)
^{2}\right\rangle =i/4$. So the value of $\xi $ is estimated by $\xi \sim
4/N\sum_{i=0}^{N}\binom{N}{i}p^{i}\left( 1-p\right) ^{N-i}\{i/2+[\left(
N-i\right) \left( N-i+2\right) /4-i/4]\}-1=\left( 1-p\right) N+1-p^{2}$, we
can thus experimentally prove a significant entanglement depth of $\left(
1-p\right) N$ qubits by measuring $\xi $.

The detection criterion in Eq. (10) is more sensitive to the bit-flip error
as this type of error significantly increases $\left\langle \left( \Delta
J_{z}\right) ^{2}\right\rangle $. With a bit flip error rate $p_{b}$ for
each qubit, the variance of $J_{z}$ is estimated by $\left\langle \left(
\Delta J_{z}\right) ^{2}\right\rangle \sim Np(1-p)$. We need $Np(1-p)<1/4$
to minimize change to $\xi $. For tens of qubits, we can tolerate bit-flip
error rate at a percent level to keep the qubits in a genuine multipartite
entangled state. Alternatively, in the limit of large $N$ with $Np(1-p)\gg
1/4$, the value of $\xi $ is estimated by $\xi \approx 1/\left[ 4p(1-p)%
\right] -1$. With a percent of bit flip error rate for each qubit, we can
experimentally prove an entanglement depth of more than $20$ qubits by
measuring $\xi $.

The criterion 1 is most appropriate for detection of the entanglement depth
in the vicinity of the Dicke state $\left\vert N/2,0\right\rangle $. It
becomes weaker for other Dicke states $\left\vert N/2,n\right\rangle $ with
increasing $\left\vert n\right\vert $. For the state $\left\vert
N/2,n/2\right\rangle $, the moments of $J_{x}$ and $J_{y}$ are bounded by $%
\left\langle J_{x}^{2}\right\rangle +\left\langle J_{y}^{2}\right\rangle
=\left\langle \mathbf{J}^{2}\right\rangle -\left\langle
J_{z}^{2}\right\rangle =N\left( N+2\right) /4-n^{2}/4$. The criterion 1 does
not take into account this bound due to a finite $\left\langle
J_{z}\right\rangle $. To derive a stronger detection criterion for the Dicke
states $\left\vert N/2,n/2\right\rangle $, we start from Eq. (6) and a
similar bound for $\left\langle J_{y}^{2}\right\rangle $. When we add up the
inequalities for $\left\langle J_{x}^{2}\right\rangle $ and $\left\langle
J_{y}^{2}\right\rangle $ both in the form of Eq. (6), we want to find a
better bound for $\left\langle \left( \Delta J_{xi}\right) ^{2}\right\rangle
_{\mu }+\left\langle \left( \Delta J_{yi}\right) ^{2}\right\rangle _{\mu }$
under a finite $\left\langle J_{z}\right\rangle $. Using the relation $%
\left\langle \left( \Delta J_{xi}\right) ^{2}\right\rangle _{\mu
}+\left\langle \left( \Delta J_{yi}\right) ^{2}\right\rangle _{\mu }\leq
\left\langle \mathbf{J}_{i}^{2}\right\rangle _{\mu }-\left\langle
J_{zi}^{2}\right\rangle _{\mu }$ and $\left\langle J_{z}^{2}\right\rangle
_{\mu }=\left\langle \left( \sum_{i=1}^{k}J_{zi}\right) ^{2}\right\rangle
_{\mu }\leq k\sum_{i}\left\langle J_{zi}^{2}\right\rangle _{\mu }$, we obtain%
\begin{eqnarray}
\left\langle J_{x}^{2}\right\rangle +\left\langle J_{y}^{2}\right\rangle
&\leq &\sum_{\mu }p_{\mu }\left[ 1+4\left\langle \left( \Delta J_{z}\right)
^{2}\right\rangle _{\mu }\right]   \notag \\
&\times &\left[ \sum_{i}m_{i\mu }(m_{i\mu }+2)/4-\left\langle
J_{z}^{2}\right\rangle _{\mu }/k\right] .  \label{11}
\end{eqnarray}%
To bound the right side of Eq. (11), we consider the two-fold average $%
\sum_{\mu }p_{\mu }\left\langle \left( \Delta J_{z}\right) ^{2}\right\rangle
_{\mu }\left\langle J_{z}^{2}\right\rangle _{\mu }=\left\langle \left\langle
\left( \Delta J_{z}\right) ^{2}\right\rangle _{\mu }\left\langle
J_{z}^{2}\right\rangle _{\mu }\right\rangle $, where $\left\langle \cdots
\right\rangle $ denotes the average over $\mu $ with the weight function $%
p_{\mu }$. For any two variables $A$ and $B$, we know their average
satisfies the following property:%
\begin{eqnarray}
\left\langle AB\right\rangle \left\langle AB\right\rangle  &=&\left\langle
A\right\rangle \left\langle B\right\rangle +\left\langle \Delta A\Delta
B\right\rangle   \notag \\
&\geq &\left\langle A\right\rangle \left\langle B\right\rangle -\sqrt{%
\left\langle \left( \Delta A\right) ^{2}\right\rangle \left\langle \left(
\Delta B\right) ^{2}\right\rangle }.  \label{12'}
\end{eqnarray}%
Taking $A$ and $B$ as $\left\langle J_{z}^{2}\right\rangle _{\mu }$ and $%
\left\langle \left( \Delta J_{z}\right) ^{2}\right\rangle _{\mu }$,
respectively, we have%
\begin{eqnarray}
-\left\langle \left\langle \left( \Delta J_{z}\right) ^{2}\right\rangle
_{\mu }\left\langle J_{z}^{2}\right\rangle _{\mu }\right\rangle  &\leq
&-\left\langle J_{z}^{2}\right\rangle \left\langle \left\langle \left(
\Delta J_{z}\right) ^{2}\right\rangle _{\mu }\right\rangle   \notag \\
&+&\left\langle \left( \Delta J_{z}^{2}\right) ^{2}\right\rangle \left(
1+2\alpha \right) ,  \label{13'}
\end{eqnarray}%
where $\left\langle \left( \Delta J_{z}^{2}\right) ^{2}\right\rangle \equiv
\left\langle J_{z}^{4}\right\rangle -\left\langle J_{z}^{2}\right\rangle ^{2}
$ and
\begin{equation}
\alpha \equiv \sqrt{\left( \left\langle J_{z}^{4}\right\rangle -\left\langle
J_{z}\right\rangle ^{4}\right) /\left( \left\langle J_{z}^{4}\right\rangle
-\left\langle J_{z}^{2}\right\rangle ^{2}\right) },  \label{14'}
\end{equation}%
which is typically close to $1$. In deriving Eq. (13), we have used $%
\left\langle \left\langle J_{z}^{2}\right\rangle _{\mu }^{2}\right\rangle
\leq \left\langle \left\langle J_{z}^{4}\right\rangle _{\mu }\right\rangle
=\left\langle J_{z}^{4}\right\rangle $ and

\begin{eqnarray}
&&\left\langle \left\langle \left( \Delta J_{z}\right) ^{2}\right\rangle
_{\mu }^{2}\right\rangle -\left\langle \left\langle \left( \Delta
J_{z}\right) ^{2}\right\rangle _{\mu }\right\rangle ^{2}  \notag \\
&=&\left\langle \left\langle J_{z}^{2}\right\rangle _{\mu }^{2}\right\rangle
-\left\langle J_{z}^{2}\right\rangle ^{2}-2\left[ \left\langle \left\langle
J_{z}^{2}\right\rangle _{\mu }\left\langle J_{z}\right\rangle _{\mu
}^{2}\right\rangle -\left\langle J_{z}^{2}\right\rangle \left\langle
\left\langle J_{z}\right\rangle _{\mu }^{2}\right\rangle \right]   \notag \\
&\leq &\left\langle \left( \Delta J_{z}^{2}\right) ^{2}\right\rangle +2\sqrt{%
\left\langle \left( \Delta J_{z}^{2}\right) ^{2}\right\rangle \left[
\left\langle J_{z}^{4}\right\rangle -\left\langle J_{z}\right\rangle ^{4}%
\right] }.  \label{15'}
\end{eqnarray}%
In the second line of Eq. (15), we use again the property in Eq. (12).
Substituting Eq. (13) into Eq. (11), we finally obtain the following bound
for any state in the form of Eqs. (1-2)%
\begin{eqnarray}
\left\langle J_{x}^{2}\right\rangle +\left\langle J_{y}^{2}\right\rangle
&\leq &\left[ 1+4\left\langle \left( \Delta J_{z}\right) ^{2}\right\rangle %
\right]   \notag \\
&\times& \max_{\left\{ m_{i\mu }\right\} }\left[ \sum_{i}m_{i\mu }(m_{i\mu
}+2)/4-\chi /k\right] ,  \label{16}
\end{eqnarray}%
where $\chi $ is defined by
\begin{equation}
\chi =\left\langle J_{z}^{2}\right\rangle -\left[ 1/4+\left\langle \left(
\Delta J_{z}\right) ^{2}\right\rangle \right] ^{-1}\left\langle \left(
\Delta J_{z}^{2}\right) ^{2}\right\rangle \left( 1+2\alpha \right) .
\label{17}
\end{equation}%
The parameter $\chi $ is determined experimentally by measuring the operator
$J_{z}$, and its value is basically given by the first term $\left\langle
J_{z}^{2}\right\rangle $, with small correction from the fluctuation of $%
J_{z}^{2}$ when the real state deviates from the Dicke state (the latter has
$\left\langle \left( \Delta J_{z}^{2}\right) ^{2}\right\rangle =0$).
Summarizing the result, we arrive at the following criterion

\textit{Criterion 2. We can experimentally measure the values of }$\xi $%
\textit{\ and }$\chi $\textit{\ (defined by Eqs. (10,17)) through detection
of the collective spin operator }$\mathbf{J}$\textit{. The system has
genuine }$m$\textit{-qubit entanglement if}

\begin{equation}
\xi >f\left( m,\chi \right) \equiv \frac{4}{N}\max_{\left\{ m_{i\mu
}\right\} }\left( \sum_{i=1}^{k}m_{i\mu }(m_{i\mu }+2)/4-\chi /k\right) -1,
\label{18}
\end{equation}%
\textit{where the maximum is taken under the constraint of }$m_{i\mu }\leq
m-1$\textit{\ and }$\sum_{i}m_{i\mu }=N$\textit{. }

With a known $\chi $, it is typically easy to calculate the function of $%
f\left( m,\chi \right) $. For instance, for the state $\left\vert
N/2,n/2\right\rangle $, $\chi \approx n^{2}/4$, and $f\left( m,\chi \right)
\approx m-\left( m-1\right) n^{2}/N^{2}$ for the simple case when $m-1$ divides $N$ and $\left(
m-1\right) n^{2}<2N^2$. Similar to the discussion made for the state $%
\left\vert N/2,0\right\rangle $, the entanglement detection criterion 2 is
pretty robust to noise, in particular the dephasing noise, and appropriate
for entanglement detection in the vicinity of the Dicke states $\left\vert
N/2,n/2\right\rangle $ with nonzero $n$.

In summary, we have proposed powerful detection criteria to experimentally
prove entanglement and quantify the entanglement depth for many-body systems
in the vicinity of arbitrary Dicke states. The criteria are based on simple
measurements of the collective spin operators and ready to be implemented in
future experiments.

This work was supported by the NBRPC (973 Program) 2011CBA00300
(2011CBA00302), the IARPA MUSIQC program, the ARO and the AFOSR MURI program.

\bigskip


\begin{thebibliography}{99}
\bibitem{1} C. A. Sackett et al., Nature 404,256 (2000); D. Leibfried
\textit{et al}., Nature \textbf{438}, 639 (2005); T. Monz etal., Phys. Rev.
Lett. 106, 130506 (2011); X.-C. Yao et al., arXiv:1105.6318.

\bibitem{2} H. H\"{a}ffner \textit{\ et al}., Nature \textbf{438}, 643
(2005).

\bibitem{3} K. S. Choi et al., Nature 468, 412 (2010).

\bibitem{4} W. Wieczorek et al., Phys. Rev. Lett. 103, 020504 (2009).

\bibitem{5} W. Duer, G. Vidal, and J. I. Cirac, Phys. Rev. A 62, 062314
(2000).

\bibitem{6} R. Raussendorf, D. E. Browne, and H. J. Briegel, Phys. Rev. A
68, 022312 (2003); M. Hein \textit{et al}, quant-ph/0602096.

\bibitem{7} J. K. Stockton, JM Geremia, A. C. Doherty, H. Mabuchi, Phys.
Rev. A 67, 022112 (2003).

\bibitem{8} M. Lewenstein, B. Kraus, J. I. Cirac, and P. Horodecki, Phys.
Rev. A \textbf{62}, 052310 (2000); B. Terhal, Phys. Lett. A 271, 319 (2000).

\bibitem{9} G. T\'{o}th and O. G\"{u}hne, Phys. Rev. Lett. \textbf{94},
060501 (2005).

\bibitem{10} G. Toth, C. Knapp, O. Guhne, and H. J. Briegel, Phys. Rev.
Lett. 99, 250405 (2007).

\bibitem{11} G. Toth, J. Opt. Soc. Am. B 24, 275 (2007).

\bibitem{note} For a many-body (mixed) state with $N$ qubits, it has
entanglement depth $m$ $\left( m\leq N\right) $ if we can experimentally
prove that it contains genuine $m$-qubit entanglement, see Ref. \cite{13}.

\bibitem{13} A. Sorensen, K. Molmer, Phys. Rev. Lett. 86, 4431 (2001).

\bibitem{14} L.-M. Duan and J. Kimble, Phys. Rev. Lett. 90, 253601 (2003);
L.-M. Duan, M. Lukin, J. I. Cirac, P. Zoller, Nature 414, 413-418 (2001).

\bibitem{15} M. J. Holland and K. Burnett, Phys. Rev. Lett. 71, 1355 (1993).

\bibitem{16} A. Acin, D. Bruss, M. Lewenstein, and A. Sanpera, Phys. Rev.
Lett. 87, 040401 (2001).
\end{thebibliography}
\end{document}